\documentclass{emulateapj}

\usepackage{apjfonts}
\usepackage{amsmath}
\usepackage{amssymb}
\usepackage{natbib}
\usepackage{xspace}
\usepackage{graphicx}
\usepackage{rotate} 
\usepackage{subfigure}
\usepackage{float}

\newcommand{\err}[2]{\ensuremath{^{+#1}_{-#2}}\xspace}
\newcommand{\e}[1]{\ensuremath{^{#1}}\xspace}
\newcommand{\ten}[1]{\ensuremath{\times 10^{#1}}\xspace}

\newcommand{\Msun}{\ensuremath{M_\odot}\xspace}

\newcommand{\nh}{$N_{\text{H}}$\xspace}
\newcommand{\NH}[1]{$N_{\text{H}, #1}$\xspace}
\newcommand{\NHgal}{$N_{\text{H\,Gal}}$\xspace}

\newcommand{\xspec}{\textsc{xspec}\xspace}

\newcommand{\etal}{et al.\ }
\newcommand{\ka}{K$\alpha$\xspace}
\newcommand{\kb}{K$\beta$\xspace}

\newcommand{\chidof}{$\chi^{2}/$dof\xspace}
\newcommand{\dchidof}{$\Delta \chi^{2}$/dof}

\newcommand{\suzaku}{\textsl{Suzaku}\xspace}

\newcommand{\xmm}{\textsl{XMM-Newton}\xspace}
\newcommand{\chandra}{\textsl{Chandra}\xspace}
\newcommand{\fluxunits}{erg\,cm$^{-2}$\,s$^{-1}$\xspace}

\shorttitle{Tracking the Complex Absorption in NGC~2110 with Two \textsl{Suzaku} Observations}
\shortauthors{Rivers et al.}

\begin{document}

\title{Tracking the Complex Absorption in NGC~2110 with Two \textsl{Suzaku} Observations} 

\author{Elizabeth~Rivers\altaffilmark{1,2}, Alex~Markowitz\altaffilmark{1,3,4}, Richard~Rothschild\altaffilmark{1}, 
Aya~Bamba\altaffilmark{5}, 
Yasushi~Fukazawa\altaffilmark{6}, 
Takashi Okajima\altaffilmark{7},
James~Reeves\altaffilmark{8}, 
Yuichi~Terashima\altaffilmark{9},
Yoshihiro~Ueda\altaffilmark{10}}
\altaffiltext{1}{University of California, San Diego, Center for Astrophysics and Space Sciences, 9500 Gilman Dr., La Jolla, CA 92093-0424, USA} 
\altaffiltext{2}{Cahill Center for Astronomy and Astrophysics, California Institute of Technology, Pasadena, CA 91125, USA} 
\altaffiltext{3}{Dr.\ Karl Remeis Sternwarte, Sternwartstrasse 7, 96049 Bamberg, Germany} 
\altaffiltext{4}{Alexander von Humboldt Fellow} 
\altaffiltext{5}{Department of Physics and Mathematics, Aoyama Gakuin University, 5-10-1 Fuchinobe, Chuo-ku, Sagamihara, Kanagawa 252-5258, Japan}
\altaffiltext{6}{Department of Physical Science, Hiroshima University, 1-3-1 Kagamiyama, Higashi-Hiroshima, Hiroshima 739-8526, Japan}
\altaffiltext{7}{NASA/Goddard Space Flight Center, Greenbelt, MD 20771, USA} 
\altaffiltext{8}{Astrophysics Group, School of Physical and Geographical Sciences, Keele University, Keele, Staffordshire, ST5 5BG, United Kingdom}
\altaffiltext{9}{Department of Physics, Ehime University 2-5 Bunkyo-cho, Matsuyama, Ehime 790-8577, Japan
\altaffiltext{10}{Department of Astronomy, Kyoto University, Kyoto 606-8502, Japan}
}
\email{erivers@caltech.edu}

\begin{abstract}
We present spectral analysis of two \suzaku observations of the Seyfert 2 galaxy, NGC~2110.  
This source has been known to show complex, variable absorption
which we study in depth by analyzing these two observations set seven years apart and by comparing to previously 
analyzed observations with the \xmm and \chandra observatories.
We find that there is a relatively stable, full-covering absorber with a column density of $\sim 3$\ten{22}~cm$^{-2}$, 
with an additional patchy absorber that is likely variable in both column density and
covering fraction over timescales of years, consistent with clouds in a patchy torus or in the broad line region.
We model a soft emission line complex, likely arising from ionized plasma and consistent with previous studies.
We find no evidence for reflection from an accretion disk in this source with no contribution from relativistically 
broadened Fe \ka line emission nor from a Compton reflection hump.

\end{abstract}

\keywords{Galaxies: active -- X-rays: spectra -- Galaxies: Individual: NGC~2110}

\section{Introduction}

X-ray spectroscopy is a valuable tool for studying the nature and geometry of material in and around active galactic nuclei (AGN).
The underlying power-law continuum is believed to arise in a hot corona very near the central supermassive black hole.
Measuring the slope (photon index) and cutoff energy of this continuum can constrain the temperature of the corona.
The Fe \ka emission line at 6.4 keV is a key diagnostic that is present in nearly all Seyfert AGN. 
Its width can indicate the distance from the supermassive black hole of the material in which the line is produced
and its energy can indicate an origin in neutral or ionized material.
Hard X-ray spectra allow study of the Compton reflection hump (see e.g., George \& Fabian 1991), a broad emission feature around 20--30 keV
which arises only from Compton-thick material (likely the geometrically thin accretion disk or the geometrically thick infrared torus).
Soft X-rays can be absorbed by material in the line of sight to the nucleus.
Type 2 Seyferts in particular tend to show significant obscuration in the soft X-ray band.
Knowing how much material is in the line of sight and the level of ionization can lead to quantifying 
how much material surrounds the central black hole and at what distance it lies.

Absorption in Seyfert AGN has been observed in a variety of forms.  Absorbers can be characterized as cold or warm (ionized), 
and may be full- or partial-covering.  Warm absorbers imprint both discrete absorption features, best observed 
in the X-rays via gratings, and broader absorption features due to multiple edges in the X-ray band 
(see e.g., Blustin \etal 2005, Behar \etal 2003).  Partial-covering absorbers due to patchy material can be complicated to model 
since they introduce degeneracies with other modeled broadband components, and can hinder proper determination of the continuum when photon 
statistics are insufficient and/or the spectral bandpass is too narrow.  Modeling partial-covering absorbers can even reduce
the (modeled) presence of relativistically-broadened Fe \ka emission lines (Miller \etal 2008; Patrick
\etal 2011).  Only with sufficiently broad X-ray coverage can one properly model all the broadband components.

An additional complication is the observation of time-variable absorption in both Seyfert 1's and 2's on a wide
range of time scales (e.g., hours--days: NGC 1365, Risaliti \etal 2009, Maiolino \etal 2010; months: Rivers \etal 2011; years: NGC 3516, Turner \etal 2011).
In some sources this variability has been modeled as absorption by discrete clumps/clouds over a wide range of length scales 
(gas commensurate with BLR clouds: NGC 3227, Lamer \etal 2003; torus-scale: Cen~A, Rivers \etal 2011).  Such evidence
suggests that accretion flows at many scales may be clumpy or filamentary, rather than uniformly smooth.  Additional observational
evidence is needed to further constrain the newest generation of models incorporating sub-pc scale and/or clumpy absorbers, including
those where the total column density along the line of sight is a viewing-angle-dependent probability (Nenkova \etal 2008).

NGC~2110 is one of the brightest Seyferts in the hard X-ray band to exhibit evidence for partial-covering absorbers.
Early observations of NGC~2110 showed significant absorption in the line of sight but with some soft emission below $\sim\,$2 keV,
either an extra component associated with the AGN such as leaked/scattered power-law continuum emission, 
or contamination from spatially extended plasma.  
While the amount of gross absorption in NGC~2110 has shown significant variability, a factor of three in ten years 
as seen by {\it HEAO 1}, {\it Ginga}, and {\it ASCA} (Risaliti \etal 2002), recent high quality X-ray spectroscopy 
has revealed that, in all likelihood, multiple absorption regions exist in this source.
Evans \etal (2007) analyzed \chandra and \xmm data, which showed that a simple absorbed power law plus leaked
emission was not sufficient to model their spectra.  Instead they found that three partial-covering absorbers 
were a much better description of their data; however due to their lack of coverage above 10 keV they were unable to 
quantify the Compton reflection hump and therefore could not be completely certain of their continuum modeling.  
Additionally, detections of lines from ionized species of Fe, Si and O in this source indicated the presence of ionized plasma in the vicinity of the AGN.
This could be in the form of a warm absorber, reflection off the ionized inner regions of an accretion disk, or emission from extended ionized plasma.

In order to characterize the soft emission and  observe variability in the absorbing complex in NGC~2110,
we have analyzed two observations with the \suzaku X-ray observatory, one from 2005 and one seven years later in 2012.
In this paper we present the results of this analysis.  Section 2 contains details of the data reduction and analysis; 
Section 3 results of the spectral fitting; and Section 4 a discussion of our conclusions.


\section{Data Reduction and Analysis}\label{sec:analysis}

\textsl{Suzaku} has two pointed instruments, the X-ray Imaging Spectrometer (XIS; Koyama et al.\ 2007) and the Hard X-ray Detector (HXD; Takahashi et al.\ 2007).
Data were taken beginning 2005-09-16 (OBSID 100024010) and 2012-08-31 (OBSID 707034010).
Data were processed with versions 2.1.6.14 (2005) and 2.8.16.34 (2012) of the \textsl{Suzaku} pipeline and recommended screening criteria were applied 
(see the \textsl{Suzaku} Data Reduction Guide\footnote{http://heasarc.gsfc.nasa.gov/docs/suzaku/analysis/abc/abc.html} for details).  
All extractions and analysis were done utilizing HEASOFT v.6.13 and \xspec v.12.6.0 (Arnaud et al.\ 1996).


\subsection{XIS Reduction}

The XIS is comprised of 4 CCD's, however XIS2 has been inoperative since 2005 November, when it was likely hit with a 
micrometeorite (see the \textsl{Suzaku} Data Reduction Guide for details).  Three of the CCD's (XIS0, XIS2 and XIS3) are front-illuminated, 
maximizing the effective area of the detectors in the Fe K bandpass, while the fourth CCD (XIS1) is back-illuminated (BI), 
increasing its effective area in the soft X-ray band ($\lesssim$\,2 keV).  

The XIS events data were taken in 3$\times$3 and 5$\times$5 editing modes, which were cleaned and summed to 
create image files for each XIS.  We extracted lightcurves and spectra from a 3 arcsec source region and four 1.5 arcsec background regions.  
After screening, the good exposure time was 101 ks per XIS for the 2005 observation and 103 ks per XIS for the 2012 observation.
The lightcurves did not show significant variability over the course of either observation.
We used the FTOOLS XISRMFGEN and XISSIMARFGEN to create the response matrix and ancillary response files, respectively.

Data were ignored above 10 keV where the effective area of the XIS begins to decrease significantly and 
below 0.5 keV due to time-dependent calibration issues of the instrumental O K edge (Ishisaki \etal 2007).
Data were ignored in the ranges 1.5--2.0 keV and 2.3--2.5 keV where there are large calibration uncertainties for the Si K complex 
and Au M edge arising from the detector's mirror system. These issues are not fully understood at the time of this writing.

\subsection{HXD Reduction}

The HXD is comprised of two detectors, the PIN diodes (12--70 keV) and the GSO scintillators (50--600 keV).
The PIN and GSO are non-imaging instruments with a 34$\arcmin$ square field of view below 100 keV and 4.5\degr ~square field of view above 100 keV.  
The HXD instrument team provides non-X-ray background event files using the calibrated GSO data for the particle 
background monitor (``tuned background''), yielding instrument backgrounds with $\lesssim$\,1.5\% systematic uncertainty at the 1$\sigma$\ 
level (Fukuzawa et al. 2009).  We simulated the Cosmic X-ray Background in \xspec using the form of Boldt (1987).

We excluded PIN data below 13 keV for the 2005 observation and 16 keV for the 2012 observation due to thermal noise (Kokubun \etal 2007).  
GSO data were usable for the 2005 observation only, since the background was lowest during the early parts of the mission, and were included up to 150 keV.
Net spectra were extracted and deadtime-corrected for a net exposure times of 81 ks per instrument for the 2005 observation and 96 ks for the 2012 observation.  


\begin{deluxetable*}{l|ccccccc}
   \tablecaption{Broadband Model Parameters \label{tabpar}}
   \tablecolumns{8}
   \startdata
\hline
\hline\\[-1mm]
Model									& \multicolumn{2}{c}{Two Absorbers} 	& \multicolumn{2}{c}{Three Absorbers} & \multicolumn{2}{c}{Two Absorbers + Soft Emission}	&	Simultaneous \\[1mm] 
Observation 								&	2005			&	2012			&	2005		&	2012			&	2005			&	2012			&	2005 / 2012\\[1mm] 
\hline\\[-1mm]
\textbf{Power Law} \\[0.5mm]
~~~ $\Gamma$    							&  1.647$\pm$0.005	& 1.679$\pm$0.005	&  1.639$\pm$0.003	&  1.675$\pm$0.004	&  1.637$\pm$0.003	& 1.678$\pm$0.004	& 1.658$\pm$0.005	\\[0.5mm]
~~~ F$_{\rm 2-10 \,keV}$\tablenotemark{1} &  143.3$\pm$0.3 & 177.7$\pm$0.4 &  143.4$\pm$0.2	&  177.0$\pm$0.3	&  143.3$\pm$0.3	& 177.8$\pm$0.4	& 143.9$\pm$0.2 / 175.4$\pm$0.7		\\[0.5mm]
\textbf{Fe \ka Line} \\[0.5mm]
~~~ $E_{\rm Fe}$ (keV)						& 6.406$\pm$0.006	& 6.379$\pm$0.008	&  6.409$\pm$0.006	&  6.378$\pm$0.008	&  6.406$\pm$0.006	& 6.378$\pm$0.008	& 6.399$\pm$0.006	\\[0.5mm]
~~~ $\sigma_{\rm Fe}$ (eV) 					&  38\err{14}{9}		& 32 ($<$50)		&  38\err{11}{9}		&  45\err{12}{22}	&  45\err{13}{8}		& 32\err{19}{27}	& 32\err{5}{14}		\\[0.5mm]
~~~ $I_{\rm Fe}$ (10$^{-5}$ ph\,cm$^{-2}$\,s$^{-1}$) & 7.1$\pm$0.4	& 9.7$\pm$0.6		&  7.4$\pm$0.5		&  9.9$\pm$0.7 	&  7.2$\pm$0.6		& 9.2$\pm$0.6		& 7.3\err{0.3}{0.6}	\\[0.5mm]
~~~ $EW$ (eV)								&   46$\pm$3		& 50$\pm$3		&  47$\pm$3		&  52$\pm$4		&  50$\pm$4		& 53$\pm$3		&		\\[0.5mm]
\textbf{Absorbers} \\[0.5mm]
~~~ \NH{1} ($10^{22}$ cm$^{-2}$)   				&  4.35$\pm$0.05	& 5.74$\pm$0.07	&  4.77$\pm$0.11	&  5.23$\pm$0.08		&  4.42$\pm$0.05	& 5.75$\pm$0.06	& 4.57$\pm$0.06 / 5.27$\pm$0.08	\\[0.5mm]
~~~ $f_1$    								&  0.745$\pm$0.003	& 0.752$\pm$0.003	&  0.47$\pm$0.01	&  0.66$\pm$0.01	&  0.753$\pm$0.003	& 0.76$\pm$0.03	& 0.785$\pm$0.003 / 0.834$\pm$0.004	\\[1.5mm]
~~~ \NH{2} ($10^{22}$ cm$^{-2}$) 	  			&  2.99$\pm$0.01	& 3.22$\pm$0.02	&  3.96$\pm$0.02	&  3.91$\pm$0.03	&  2.97$\pm$0.02	& 3.23\err{0.02}{0.12} & 2.82$\pm$0.02		\\[0.5mm]
~~~ $f_2$    								&  1.0 ($\geq$ 0.99)	&  1.0 ($\geq$ 0.99)	&  0.98$\pm$0.01	& 0.98$\pm$0.01 	&   1.0 ($\geq$ 0.99)	&  1.0 ($\geq$ 0.99)	& 1.0*	\\[1.5mm]
~~~ \NH{3} ($10^{22}$ cm$^{-2}$)    			&				&				&  0.27$\pm$0.01	&  0.26$\pm$0.02	&				&				&		\\[0.5mm]
~~~ $f_3$    								&				&				& 1.0*			&	1.0*			&				&				&		\\[0.5mm]
\textbf{Soft Emission (Gaussian)} \\[0.5mm]
~~~ $E_{\rm SX}$ (keV)						&				&				&				&				&  0.84$\pm$0.02	& 0.89$\pm$0.02	& 0.89$\pm$0.01	\\[0.5mm]	
~~~ $\sigma_{\rm SX}$ (eV)					&				&				&				&				&  190\err{30}{10}	& 130\err{40}{10}	& 100$\pm$10		\\[0.5mm]
~~~ $I_{\rm SX}$ (10$^{-5}$ ph\,cm$^{-2}$\,s$^{-1}$) &				&				&				&				&  12.2\err{2.1}{0.5}	& 9.2\err{2.3}{0.7}	& 4.8	$\pm$0.4		\\[0.5mm]
${\bf \chi}^{\bf 2}$/\textbf{dof}					&   1299/694		&	1082/520		&   912/693		&	706/519		&  882/691		& 671/517			& 1674/1197	\\[-1mm]
\enddata
\tablecomments{Best fit parameters for three broadband models including a power-law continuum and Gaussian Fe \ka line in all models.  The soft emission complex is modeled with a phenomenological Gaussian in the Soft Emission Model.  The power law flux is unabsorbed.  Note that the width of the soft emission Gaussian ($\sigma_{\rm SX}$) does not signify physical broadening but is rather the blend of multiple soft emission lines.  The "*" symbol indicates a frozen parameter.}
\tablenotetext{1}{2--10 keV flux is given in units of 10\e{-12} erg cm\e{-2} s\e{-1}}
\end{deluxetable*}


\begin{deluxetable}{lcccc}
   \tablecaption{Soft Emission Components \label{tabsx}}
   \tablecolumns{5}
   \startdata
\hline
\hline\\[-1mm]
\textbf{Collisionally }		&	\multicolumn{2}{c}{$kT$ (eV)}	&  \multicolumn{2}{c}{Norm (10\e{-5})}		\\[0.5mm]
\textbf{Ionized}			&	2005		     &	  2012		&	2005			&	2012				\\[0.5mm]
\hline\\[-1mm]
\textsc{APEC}  				&  910\err{40}{20} &	  970$\pm$40	&	6.6\err{0.6}{1.0} &	11.5\err{1.6}{1.1}	\\[0.5mm]
\hline
\hline\\[-1mm]
\textbf{Photoionized}		&	$E$ 		&	Width ($\sigma$/$kT$) 	&	\multicolumn{2}{c}{Norm (10\e{-5} photons cm\e{-2} s\e{-1})}	\\[0.5mm]
					&	(keV)	&	(eV)		&	2005		&	2012				\\[0.5mm]
\hline\\[-1mm]
O \textsc{VII}			&	0.56*	&	1*		&	1.6$\pm$1.0	&	$\leq$7.1		\\[0.5mm]
O \textsc{VIII}			&	0.65*	&	1*		&	1.6$\pm$0.4	&	$\leq$2.5		\\[0.5mm]
Fe L					&	0.72*	&	1*		&	1.1$\pm$0.3	&	1.2$\pm$0.9	\\[0.5mm]
O \textsc{VII} RRC		&	0.74*	&	100*		&	2.7$\pm$0.4	&	1.3$\pm$0.8	\\[0.5mm]
Fe L					&	0.84*	&	1*		&	0.8$\pm$0.2	&	1.2$\pm$0.5	\\[0.5mm]
O \textsc{VIII} RRC		&	0.87*	&	100*		&	3.4$\pm$0.3	&	7.34$\pm$0.5	\\[-1.5mm]
\enddata
\tablecomments{Best fit parameters for the two absorber fit plus a collisionally ionized plasma (\textsc{APEC} ) or a photoionized plasma (complex of lines and RRC features).  Width corresponds to $\sigma$ for the Gaussian lines and to $kT$ for the RRC features (\textsc{REDGE} component in \xspec).  The "*" symbol indicates a frozen parameter.}
\end{deluxetable}

\section{Spectral Fitting}

All spectral fitting was done in XSPEC utilizing solar abundances of Wilms \etal (2000) and cross-sections from Verner \etal (1996). 
All fits included absorption by a Galactic column with \NHgal=\,1.62$\times 10^{21}$\,cm$^{-2}$ (Kalberla \etal 2005).
Uncertainties are listed at the 90\% confidence level ($\Delta \chi^2$ = 2.71 for one interesting parameter).

\subsection{The Hard X-ray Bandpass}

We began by fitting our data from each observation epoch separately above 3 keV, 
including the XIS+PIN+GSO for the 2005 observation in the range 3--150 keV and the XIS+PIN for the 2012 observation in the range 3--70 keV.
For the 2012 observation we froze the PIN normalization relative to XIS0 at 1.16, the calibrated value (Kokubun \etal 2007).  
For the 2005 observation, however, this led to poor fit statistics, likely because the background files were generated during the earliest phase of the mission.
Therefore we left the PIN normalization as a free parameter for the 2005 observation only, generally getting values of $\sim\,$1.05.
Normalization constants (relative to XIS0) were left free for all other XIS's and were very close to 1.

Each observation was well-fit by a simple absorbed power law plus a narrow Gaussian line around 6.4 keV to model the Fe \ka emission line (for more details on the Fe K bandpass see Section 3.3).  
We found best fit parameters of $\Gamma = 1.67$, \nh = 6.97 $\times 10^{22}$ cm$^{-2}$ and \chidof = 471/353.
Fitting a reflection spectrum to this interval using the \textsc{pexrav} model in \xspec did not yield an improvement in the fit.
We found an upper limit to Compton reflection of $R \lesssim 0.1$, assuming an inclination of 60\degr. 
Testing for a high energy cutoff in the spectrum with the \xspec model \textsc{highecut} yielded a lower limit of $E_{\rm cut} \gtrsim 250$ keV.

\subsection{Broadband Modeling}

\begin{figure}
  \plotone{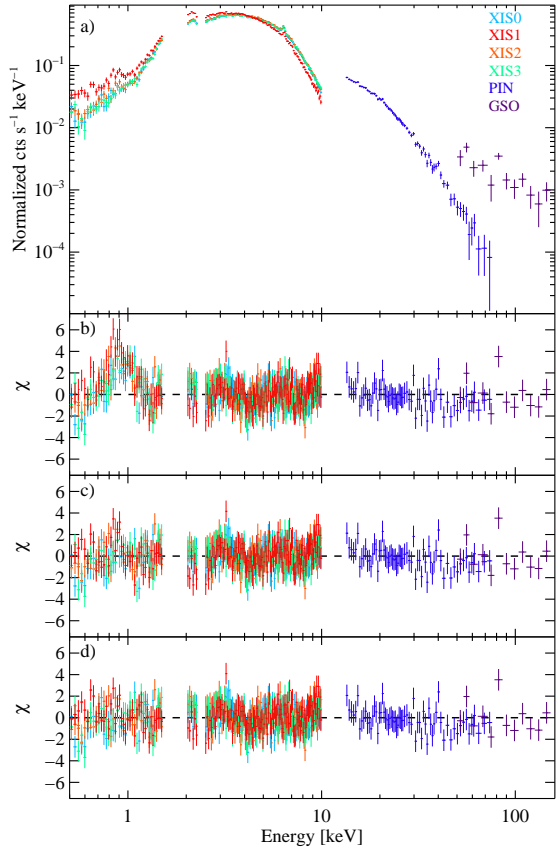}
  \caption{\suzaku XIS and HXD data for the 2005 observation.  Panel a) shows the data; b) residuals to the Two-Absorber model; c) residuals to the Three Absorber model; d) residuals to the Soft Emission model.  Parameters are listed in Table \ref{tabpar}.  Note that residuals around 8--10 keV are due to calibration uncertainties which are not well understood at the time of this writing.}
  \label{figspec1}
\end{figure}

\begin{figure}
  \plotone{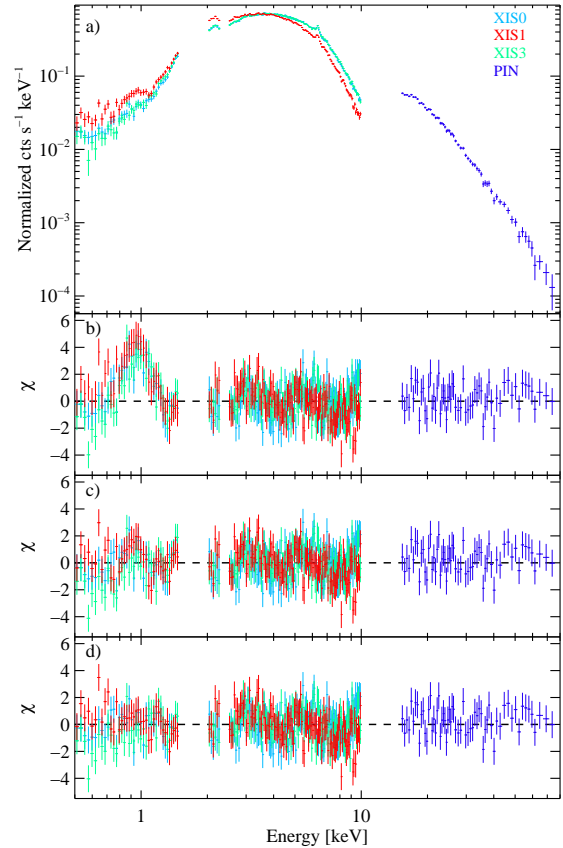}
  \caption{\suzaku XIS and HXD data for the 2012 observation.  Panel a) shows the data; b) residuals to the Two-Absorber model; c) residuals to the Three Absorber model; d) residuals to the soft emission (Gaussian) model.  Parameters are listed in Table \ref{tabpar}. Note that residuals around 8--10 keV are due to calibration uncertainties which are not well understood at the time of this writing.}
  \label{figspec7}
\end{figure}

Including data from 0.4--3 keV in our fits (excluding the intervals 1.5--2.0 keV and 2.3--2.5 keV) we noticed obvious residuals below $\sim$\,3 keV
due to the complex absorbers that have been seen previously in this source (Guainazzi \& Bianchi 2007; Evans \etal 2007). 
We tried three models for the broadband data shown in Figures \ref{figspec1} and \ref{figspec7}; parameters are given in Table \ref{tabpar}.
As an initial baseline, we first tried fitting two zones of absorption, allowing column density and covering fraction to be free for both layers. 
This "Two-Absorber" model had best-fit values of $N_{\rm H,1}$ near 4.4 and 5.7 $\times 10^{22}$ cm$^{-2}$ (2005 and 2012 respectively), 
and covering fractions $f_1$ near 75$\%$, as listed in Table~1. 
$N_{\rm H,2}$ was close to $3 \times 10^{22}$ cm$^{-2}$ with a covering fraction of 1.0 (fully covering). 
This model did not provide a satisfactory fit and left strong, positive data/model residuals below 2 keV, as illustrated in Figures 1b and 2b.

Next, we applied the best-fit model of Evans et al.\ (2007) which included a third zone of absorption with a fixed covering fraction of 1 (layer 3 in the Three-Absorber model in Table ~1).
The fit was a significant improvement over previous fits (reduced $\chi^2$ values of 1.32 and 1.36).
$N_{\rm H,1}$ and $N_{\rm H,2}$ remained at values similar to the Two-Absorber model. 
The covering fraction for layer 2, $f_2$, remained very close to unity, with $\sim$2\% leaked emission to model the positive residuals below 2 keV. 
However, as shown in Figures 1c and 2c, the residuals below 2 keV were still not modeled optimally.

Finally, we tried adding a phenomenological Gaussian component centered at $\sim\,$0.9 keV to the Two-Absorber model to model a blend of soft X-ray emission lines.
This  ``Two Absorbers plus Soft Emission" model (henceforth the "Soft Emission" model) fit the broadband data much better than the previous models, 
both in terms of data/model residuals (Figures 1d and 2d) and reduced $\chi^2$. Best-fit parameters are listed in Table~1;
parameters for layers 1 and 2 were similar to those in the Two-Absorber model.  Because we are using one component to model a
blend of emission lines, the width $\sigma$ of the Gaussian component is not physically meaningful.

We also tried modeling warm absorption using an XSTAR table in \xspec with an additional cold full-covering absorber, 
since absorption by highly ionized gas can mimic soft X-ray emission lines.
However the fit was only a mild improvement on the Two-Absorber model and failed to model the soft emission with very poor residuals below 2 keV.  

Guainazzi \& Bianchi (2007) found evidence in this source for a radiative recombination continuum (RRC) feature due to  O \textsc{VIII} at 0.87 keV, 
which was tentatively confirmed by Evans \etal (2007), who also found hints of a Ly$\alpha$ emission line from O \textsc{VIII} at 0.65 keV.  
These features would be present if the extended plasma around the source were photoionized, 
and would likely be accompanied by other lines at low energies such as from O \textsc{VII} and Fe L transitions.
We have therefore modeled a physically motivated soft emission complex due to photoionized plasma with a complex of lines and RRC features.
Parameters for this model are listed in Table \ref{tabsx}.  Note that the absorbers in this model  (with the exception of the Galactic column) are applied to the power law only.  
Evans \etal (2007) ruled out a single-temperature \textsc{APEC} component and reasoned that while collisionally ionized plasma was likely present, 
photoionization must also be a significant process in this source.
\chidof values for the photoionized complex versus \textsc{APEC}  were 976/710 versus 983/713 for the 2005 observation and 745/544 versus 739/543 for the 2012 observation.
Parameters for the Two-Absorber model plus an \textsc{APEC}  component are also given in Table \ref{tabsx}.


\subsection{The Fe K Bandpass}

We analyzed the Fe K bandpass in more detail only after obtaining satisfactory models for the broadband continuum.
We have detected a weak, narrow Fe \ka line with a width of around 20--60 eV, consistent with measurements made by Evans \etal (2007) using \chandra data.  
Fe \kb emission was undetectable with an upper limit to the intensity of 7.2$\times$10$^{-5}$ ph\,cm$^{-2}$\,s$^{-1}$ and $EW \lesssim 4$ eV
(width tied to that of the Fe \ka line and energy centroid frozen at 7.056 kEV). 
We found no evidence for emission from ionized Fe with upper limits of $EW \lesssim 4$ eV for Fe {\sc XXV} and $EW \lesssim 6$ eV for Fe {\sc XXVI}  
(widths tied to that of the \ka line and energy centroids frozen at 6.70 keV and 6.97 keV, respectively).  

We also tested for relativistically broadened Fe \ka emission using the \textsc{diskline} model in \xspec in addition to a narrow Gaussian
with a fixed width ($\sigma$ = 1 eV).  The improvement in fit was negligible (\dchidof = $-$2) with an upper limit to the intensity of the \textsc{diskline} of 
2.4$\times$10$^{-5}$ ph\,cm$^{-2}$\,s$^{-1}$ ($EW \lesssim 15$ eV).


\subsection{Simultaneous Fitting}

\begin{figure}
  \plotone{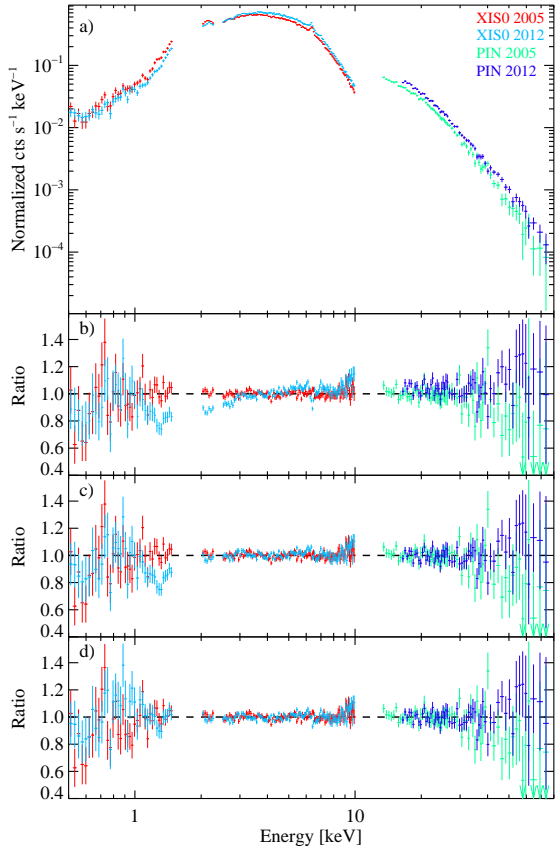}
  \caption{\suzaku XIS0 and PIN data for the 2005 and 2012 observations.   Panel a) shows the data; b) ratios for the soft emission model with only the flux of the power law and intensity of the Fe line left free; c) ratios when \NH{1} is allowed to vary between the observations; d) ratios when \NH{1} and $f_1$ are allowed to vary between the observations.  The other XIS data are not shown for clarity purposes only; all data were used in the fitting.}
  \label{figspecall}
\end{figure}

It is clear that the source flux and spectral characteristics remained fairly stable over the course of $\sim$\,6 years, however slight changes seem to have occurred.
In order to investigate this spectral variability thoroughly, we performed simultaneous fitting of both \suzaku observations.
We used the Soft Emission model from Table \ref{tabpar} with a Gaussian soft emission component, tying all parameters between 
the observations except F$_{2-10}$, $I_{\rm Fe}$, and the instrumental renormalization constants 
(Note that  for the 2012 observation the PIN renormalization constant was still tied at 1.16 relative to XIS0).  
This provided a poor fit (\chidof = 2793/1199) and it was clear that the difference in 
the levels of absorption between the two observations was real.  Therefore we untied the column density of the first absorber (\NH{1}), 
yielding an improved fit with \chidof = 2079/1198.  Untying just the covering fraction, $f_1$, but not the column density gave a slightly better \chidof value of 1943/1198.
Untying both \NH{1} and $f_1$ yielded a good fit with \chidof = 1649/1197.  The data and ratio plots for these fits are shown in Figure \ref{figspecall}.
It was not significant to allow $I_{\rm SX}$, \NH{2}, or $f_2$ to vary between the observations.



\begin{figure}
  \plotone{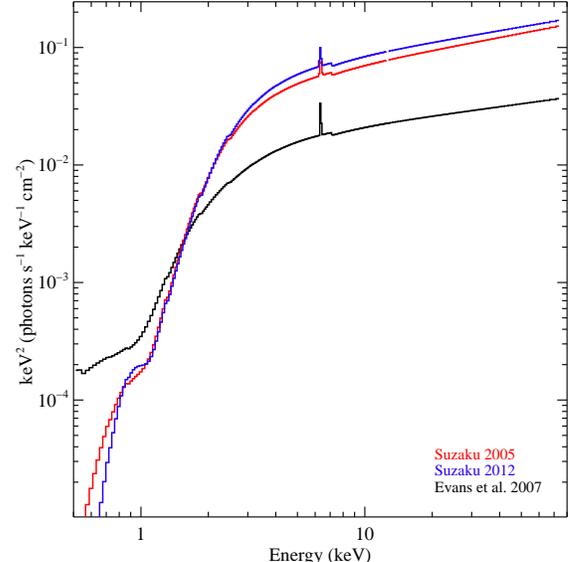}
  \caption{Energy density plot for the best-fit to the \suzaku observations as well as the best-fit three absorber model of Evans \etal (2007) for comparison.  While the continuum level is much higher in both \suzaku observations than the \xmm + \chandra observation, the soft emission is considerably lower, indicating the possibility of an additional variable leaked/scattered power law component which was too weak to be significantly detected/untangled in our observations.}
  \label{figmods}
\end{figure}

\section{Discussion and Conclusions}

With a high quality spectrum above 10 keV we are able to rule out the presence of a strong Compton reflection hump in this source.  This simplifies our continuum 
modeling and means we can compare directly with previous observations which may have neglected this component due to lack of high energy coverage.  
We adopt the Soft Emission model described in Section 3.2 with parameters given in Table \ref{tabpar} as the best-fit model for the remainder of this paper.

The 2--10 keV unabsorbed luminosity of this source is $L_{2-10} \sim 3-4 \times 10^{42}$\,erg\,s\e{-1}.  
This corresponds to a bolometric luminosity of roughly $L_{\rm Bol} \sim 10^{44}$\,erg\,s\e{-1} using a bolometric correction of $\sim$20 from Marconi \etal (2004), their Figure 3b.
Assuming a  a black hole mass of 2.5$\times 10^7$ \Msun (Merloni \etal 2003), we calculate the Eddington fraction of this source to be  $L_{\rm Bol} / L_{\rm Edd} \sim 0.03$.

\subsection{The Complex of Absorbers}

The absorbers in this source are clearly complex and time-variable.  Layer 1 had the highest column density, and we had to model it as a partial coverer to achieve a good fit.
Our observations show an increase in \NH{1} of $\sim\,$20\% in 6 years with only a slight change in the measured covering fraction.
Evans \etal (2007) analyzed \chandra-HETGS/\xmm observations from 2001/2003 and found that their data were well-fit with the Three-Absorber model with column densities and covering 
fractions of \NH{1}~=~12.8\ten{22} cm$^{-2}$ with $f_1$~=~0.32, \NH{2}~=~2.76\ten{22} cm$^{-2}$ with $f_2$~=~0.96, and \NH{3}~=~7.7\ten{20} cm$^{-2}$ with $f_3$~=~1.
However, this model did not fit the \suzaku data as well as only two absorbers with the addition of a soft emission component.
Additionally, Evans \etal (2007) found a much higher column density for layer 1 with a slightly lower covering fraction.
Figure 4 shows the energy density plots of the best-fit models to the \suzaku observations and the model of Evans \etal (2007) for comparison.

If this change in the column density and covering fraction of layer 1 is a real effect and not instrumental 
(which could be due to bandpass differences and/or splicing non-simultaneous data
as was done with the \chandra data), then the absorber underwent a significant change between 2003 and 2005, but experienced only a very slight change
between 2005 and 2012.  This could occur if a particularly dense, compact cloud was passing through the line of sight in 2001--2003, as has been observed on 
similar timescales in other Seyferts such as NGC~3516 (Turner \etal 2008), NGC 3227  (Lamer \etal 2003), and Cen~A (Rivers \etal 2011).
The level of absorption in 2005--2012 could represent a baseline level of absorption with more diffuse clouds causing the lower column 
density and higher covering fraction observed.

Fritz \etal (2006) modeled the mid-infrared emission of NGC~2110 with a dusty torus model, finding an equatorial column density of 
\nh = 2.59$\times 10^{23}$ cm\e{-2}.  Given the double sided radio jet emission in this source (Evans \etal 2006) it seems probable that the source is roughly edge-on
and should therefore be absorbed by the infrared torus, however this column density is clearly much higher than that seen in the X-rays.
This could be easily reconciled if the torus density were dependent on the polar angle and therefore on the viewing angle, 
or if the dust/gas ratio is very different from that assumed.
Additionally, some have claimed that AGN tori must be clumpy (Risaliti \etal 2002; Nenkova \etal 2008), which could account for the partial-covering nature of the first layer of absorption.

Alternately, the first layer of absorption could be quite close in to the central black hole, commensurate with the broad line region (BLR) clouds, 
as has been inferred in, for example, MCG--6-30-15 (McKernan \& Yaqoob 1998), NGC~4051 (Guainazzi \etal 1998), and NGC~1365 (Risaliti \etal 2009).  
NGC~2110 has a ``hidden'' BLR, detectable only in the infrared (Reunanen \etal 2003).

The second layer of absorption seems to have remained fairly stable over $\sim$\,11 years, with a column density of \NH{2}$\,\sim\,$3\ten{22}~cm\e{-2}
in the \chandra, \xmm, and \suzaku observations, with a covering fraction $\gtrsim$0.96.  
The stability of the column density over time would not be expected for a patchy, partial-covering absorber.
Additionally, our Soft Emission model does not require this absorber to be partial-covering.  
This leads us to one of two scenarios: either this material is very homogeneous (in which case it is likely not partial-covering), 
or it is a partial-covering cloud located far from the central black hole.
A transit of at least 11 years corresponds to a radius of $\gtrsim$\,10\e{18} cm or $\gtrsim$\,50 pc (Lamer \etal 2003: Equation 3), 
assuming an ionization parameter of 1 erg\,cm\,s\e{-1} and a black hole mass of 2.5$\times 10^7$ \Msun (Merloni \etal 2003).

Layer 2 could also be located much farther out from the nucleus, in the galaxy itself.
The inclination dependence of dust extinction for disk-dominated galaxies has been derived in general by both Driver \etal (2007) and Shao \etal (2007),
with lines of sight through even relatively edge-on disks contributing at most $\lesssim\,$2 magnitudes of optical extinction. Ê
The ratio of minor to major axes (0.78, Two Micron All Sky Survey Team's 2MASS Extended objects catalog, 2003) yields the inclination of the disk: 
tilted 39\degr away from the plane of the sky. At this inclination, the average optical extinction is typically less than about half a magnitude.
Assuming for simplicity similar B- and V-band magnitudes and a Galactic dust/gas ratio of \nh = 1.8$\times 10^{21}$ cm\e{-2} $\times$ $A_{\rm V}$ 
(Predehl \& Schmidt 1995), this corresponds to a column density of only $\sim 10^{21}$ cm\e{-2}.
Of course, we cannot rule out the possibility that the line of sight in NGC~2110 passes through an overdense/overdusty region in the host galaxy 
such as a giant molecular cloud.  Assuming this is not the case, however, it seems likely that Layer 2 resides within the nuclear region.

\subsection{The Soft Emission}

There are several possibilities for the source of the soft emission in NGC~2110.  
One thing that needs to be ruled out is contamination from nearby point sources.
The three closest point sources identified by \chandra all had flat spectra with 2--10 keV fluxes 
below 10\e{-13} \fluxunits  (Evans \etal 2006; Evans \etal 2007) and would have negligible impact on our measurements.

The presence of soft emission and tentative detections of ionized emission lines in the \chandra data (Evans \etal 2007) hinted at the possibility of an ionized reflector.  
However, the \xspec model \textsc{reflionx} was unable to reproduce the spectral shape. 
Additionally, the lack of a Compton reflection hump, broadened Fe lines, or emission from highly ionized Fe is consistent with there being no reflection from an accretion disk.
We were also able to rule out the presence of a mildly ionized warm absorber as the cause of the soft spectral shape.

The two ionized plasma models that we fit to our data could be due in part to the extended soft X-ray plasma seen by \chandra (Evans \etal 2006).
Our collisionally ionized plasma temperature of $\sim\,$910--970 eV is consistent with the Southern extended emission 
that Evans \etal (2006) fit with an \textsc{APEC}  model, finding a temperature of 960$\pm200$ eV.
The normalization for this component was a factor of $\sim\,$10 lower as found by the \chandra observations, however the \chandra extraction region was very limited and can only be taken as a lower limit to the strength of the emission.
Since an excess around 1 keV was also seen by the \chandra observation of the nucleus of NGC~2110 (which excluded the extended emission regions), 
it seems likely that there are multiple ionized regions contributing to the soft emission.  Unfortunately we are unable to disentangle them with the CCD resolution of the XIS, 
and the lack of photons picked up by the gratings of both \chandra and \xmm mean we cannot use that data to break the degeneracies, either between the 
different regions or between collisional and photo-ionization processes.

\subsection{The Fe K Emission Complex}

Early modeling showed evidence for a relativistically broadened Fe \ka~line in this source 
(Weaver \& Reynolds 1998, Turner \etal 1998), however partial-covering absorber models and detailed soft X-ray 
coverage with {\it Chandra} and {\it XMM-Newton} dramatically reduced the significance of such a component (Evans \etal 2007).  
Our data are in good agreement with Evans \etal (2007), indicating that if there is a contribution from a broadened Fe line, it is very weak.
The lack of a broadened line could indicate that the accretion disk is truncated and/or surrounds a radiatively inefficient flow (Esin et al.\ 1997).  
Another possibility is that the inner disk is too highly ionized to produce appreciable Fe line emission. 
Though the characteristic rollover above 30 keV due to the Compton reflection hump would still be present, 
we are unable to place constraints on such a component (modeled with, e.g., \textsc{xillver}; Garci\'a \etal 2013).
X-ray missions with high sensitivity in the 20--100 keV range such as \textsl{NuSTAR} or \textsl{Astro-H} may 
be able to measure the contribution of reflection from extremely ionized material.

The overall amount of Fe emission in this source is quite weak.  To calculate the expected amount of Fe emission from the absorbers detected we can use 
a thin-shell approximation and the following equation based on Yaqoob \etal (2001):

\begin{center}
\begin{equation}  EW_{\rm K\alpha} = f_{\rm c}\, \omega\, f_{\rm K\alpha}\, A_{\rm abund}\, N_\text{H}\, \frac{\int_{E_{\rm K-edge}}^{\infty} P(E)\, \sigma_{\rm ph}(E)\, {\rm d}E}{P(E_{\rm line})}  \end{equation}
\end{center}

with $f_{\rm c}$ the covering fraction of the absorber, $\omega$ the fluorescent yield, $f_{\rm K\alpha}$ the fraction of photons that go into 
producing the \ka line, $P$(E) the continuum power law, and $\sigma_{\rm ph}(E)$ the K-shell absorption cross section as a function of energy.  

Assuming solar abundances, $A_{\rm abund}$, and using values for the fluorescent yield and cross section from Yaqoob \etal (2001),  
along with our measured continuum and absorber parameters, we can calculate the contributed $EW$ for each absorber.
For the 2005 observation we find that $EW_{\rm Fe\,K\alpha} = EW_1 + EW_2 = 30 + 26 = 56$ eV.
For the 2012 observation we find that $EW_{\rm Fe\,K\alpha} = EW_1 + EW_2 = 35 + 26 = 61$ eV.
This is consistent with the measured values of 46$\pm$4 eV and 51$\pm$3 eV, respectively, allowing for variability of the source, 
and assuming that the measured covering fraction is approximately equal to the global covering fraction.

Our measured Fe \ka line width ($\sim$ 32--45 eV) corresponds to $v_{\rm FWHM}$ of 4000--5600 km s\e{-1}.  
For an assumed black hole mass of 2.5$\times 10^7$ \Msun, this corresponds to an inner radius of 0.004--0.008 pc.
If the Fe line truly is associated with the absorbers then this is also the lower limit to the inner radius of the absorbing material.
This radius is consistent with typical values for the BLR in Seyferts, hinting that the absorbers could be commensurate with the BLR clouds.
Unfortunately, the widths of the IR broad lines are not resolved in this source.

\smallskip
\subsection{Summary}

We have analyzed two \suzaku observations of the Seyfert 2 galaxy, NGC~2110.  This source has been known to have complex, 
variable absorption previously observed with the \xmm and \chandra observatories (Evans \etal 2007).
We found that there is a relatively stable full-covering absorber with a column density of $\sim\,3$\ten{22} cm$^{-2}$, 
with an additional patchy absorber that is highly variable, possibly in the form of a patchy torus or broad line region.
We found that the soft emission, which has been interpreted in several ways in the past, likely arises in an ionized plasma.  
There is likely contribution by the extended soft X-ray plasma detected by \chandra, as well as from within the nucleus. 
We tested both collisionally ionized and photoionized models, but were unable to distinguish between them.
Instruments with a combination of high effective area and spectral resolution such as those aboard the upcoming \textsl{Astro-H} mission 
(Takahashi \etal 2012) will be necessary to fully disentangle these components.  

We find no evidence for a Compton reflection hump in this source, i.e., there seems to be no evidence for the existence of any
circumnuclear gas with a column higher than 1.2 $\times 10^{23}$\,cm$^{-2}$ in NGC~2110.  We also find no evidence for a
relativistically broadened Fe K$\alpha$ line.  This may be due to a radiatively-inefficient or advection-dominated accretion flow (e.g.,
Esin \etal 1997) feeding the black hole in this source.  The lack of these X-ray spectral signatures also occurs in the radio-loud AGN
Cen~A (e.g., Markowitz \etal 2007), hinting that in at least some sources accretion may proceed in a different mode than typical Seyferts.  
Where Compton humps and broad Fe K$\alpha$ lines are not significantly detected it may be that the Compton-thick gas is poorly illuminated 
by the central source (e.g., a Compton-thick torus with a very large opening angle), the accretion is radiatively inefficient, and/or gas is not accumulating to Compton-thick levels.

\begin{acknowledgments}
This research has made use of data obtained from the \textsl{Suzaku} satellite, a collaborative 
mission between the space agencies of Japan (JAXA) and the USA (NASA).
This work has made use of HEASARC online services, supported by NASA/GSFC, and the NASA/IPAC Extragalactic Database, 
operated by JPL/California Institute of Technology under contract with NASA.
This research was supported by Grant NNX13AF33G.
\end{acknowledgments}



\end{document}